\documentclass[numberedappendix,twocolumn,tighten]{aastex62}
\usepackage{natbib}
\begin{document}

\title{1D Kinematics from stars and ionized gas at $z\sim0.8$ from the LEGA-C spectroscopic survey of massive galaxies}
\correspondingauthor{Rachel Bezanson}
\email{rachel.bezanson@pitt.edu}

\author[0000-0001-5063-8254]{Rachel Bezanson}
\affiliation{Department of Physics and Astronomy, University of Pittsburgh, Pittsburgh, PA 15260, USA}

\author[0000-0002-5027-0135]{Arjen van der Wel}
\affiliation{Sterrenkundig Observatorium, Universiteit Gent, Krijgslaan 281 S9, B-9000 Gent, Belgium}
\affiliation{Max-Planck Institut f{\"u}r Astronomie, K{\"o}nigstuhl 17, D-69117, Heidelberg, Germany}

\author[0000-0001-5937-4590]{Caroline Straatman}
\affiliation{Sterrenkundig Observatorium, Universiteit Gent, Krijgslaan 281 S9, B-9000 Gent, Belgium}

\author[0000-0003-4196-0617]{Camilla Pacifici}
\affiliation{Space Telescope Science Institute, 3700 San Martin Drive, Baltimore, MD 21218, USA}

\author[0000-0002-9665-0440]{Po-Feng Wu}
\affiliation{Max-Planck Institut f{\"u}r Astronomie, K{\"o}nigstuhl 17, D-69117, Heidelberg, Germany}

\author[0000-0001-6371-6274]{Ivana Bari\v{s}i\'c}
\affiliation{Max-Planck Institut f{\"u}r Astronomie, K{\"o}nigstuhl 17, D-69117, Heidelberg, Germany}

\author[0000-0002-5564-9873]{Eric F. Bell}
\affiliation{Department of Astronomy, University of Michigan, 1085 South University Ave., Ann Arbor, MI 48109, USA}

\author[0000-0002-1590-8551]{Charlie Conroy}
\affiliation{Harvard-Smithsonian Center for Astrophysics, Cambridge, MA 02138, USA}

\author[0000-0003-2388-8172]{Francesco D'Eugenio}
\affiliation{Sterrenkundig Observatorium, Universiteit Gent, Krijgslaan 281 S9, B-9000 Gent, Belgium}

\author[0000-0002-8871-3026]{Marijn Franx}
\affiliation{Leiden Observatory, Leiden University, P.O.Box 9513, NL-2300 AA Leiden, The Netherlands}

\author[0000-0002-9656-1800]{Anna Gallazzi}
\affiliation{INAF-Osservatorio Astrofisico di Arcetri, Largo Enrico Fermi 5, I-50125 Firenze, Italy}

\author{Josha van Houdt}
\affiliation{Max-Planck Institut f{\"u}r Astronomie, K{\"o}nigstuhl 17, D-69117, Heidelberg, Germany}

\author[0000-0003-0695-4414]{Michael V. Maseda}
\affiliation{Leiden Observatory, Leiden University, P.O.Box 9513, NL-2300 AA Leiden, The Netherlands}

\author[0000-0002-9330-9108]{Adam Muzzin}
\affiliation{Department of Physics and Astronomy, York University, 4700 Keele St., Toronto, Ontario, Canada, MJ3 1P3}

\author[0000-0003-2552-0021]{Jesse van de Sande}
\affiliation{Sydney Institute for Astronomy, School of Physics, A28, The University of Sydney, NSW, 2006, Australia}

\author[0000-0001-8823-4845]{David Sobral}
\affiliation{Department of Physics, Lancaster University, Lancaster LA1 4YB, UK}
\affiliation{Leiden Observatory, Leiden University, P.O.Box 9513, NL-2300 AA Leiden, The Netherlands}

\author[0000-0003-3256-5615]{Justin Spilker}
\affiliation{Department of Astronomy, University of Texas at Austin, 2515 Speedway, Stop C1400, Austin, TX 78712, USA}

\shortauthors{Bezanson et al.}
\shorttitle{Stellar and Ionized Gas 1D Kinematics at $z\sim0.8$}


\begin{abstract}
We present a comparison of the observed, spatially integrated stellar and ionized gas velocity dispersions of $\sim1000$ massive ($\log\,M_{\star}/M_{\odot}\gtrsim\,10.3$) galaxies in the Large Early Galaxy Astrophysics Census (LEGA-C) survey at $0.6\lesssim\,z\lesssim1.0$. The high $S/N\sim20{\rm\AA^{-1}}$ afforded by 20\,hour VLT/VIMOS spectra allows for joint modeling of the stellar continuum and emission lines in all galaxies, spanning the full range of galaxy colors and morphologies. These observed {integrated} velocity dispersions (denoted as $\sigma'_{g, int}$ and $\sigma'_{\star, int}$) are related to the intrinsic velocity dispersions of ionized gas or stars, but also include rotational motions through beam smearing and spectral extraction. We find good average agreement between observed velocity dispersions, with $\langle\log(\sigma'_{g, int}/\sigma'_{\star, int})\rangle=-0.003$. This result does not depend strongly on stellar population, structural properties, or alignment with respect to the slit. However, in all regimes we find significant scatter between $\sigma'_{g, int}$ and $\sigma'_{\star, int}$, with an overall scatter of 0.13\,dex of which 0.05\,dex is due to observational uncertainties. For an individual galaxy, the scatter between $\sigma'_{g, int}$ and $\sigma'_{\star, int}$ translates to an additional uncertainty of $\sim0.24\rm{dex}$ on dynamical mass derived from $\sigma'_{g, int}$, on top of measurement errors and uncertainties from Virial constant or size estimates. We measure the $z\sim0.8$ stellar mass Faber-Jackson relation and demonstrate that emission line widths can be used to measure scaling relations. However,  these relations will exhibit increased scatter and slopes that are artificially steepened by selecting on subsets of galaxies with progressively brighter emission lines. 
\end{abstract}

\section{Introduction}

One of the most fundamental properties of a galaxy is the depth of its gravitational well, which can be measured from integrated or spatially-resolved data based on the kinematics derived from either gas or stars. The line-of-sight ``velocity dispersion,'' or the second moment of the velocity distribution function, is the simplest measure of a galaxy's potential well that can be measured from a 1D spectrum. Observationally, one measures an intrinsic velocity dispersion ($\sigma$) that is convolved with rotational motions by the point spread function and 1D spectral extraction in addition to projection effects along the line-of-sight. Therefore throughout this paper we refer to this measured velocity width as $\sigma'_{int}$ or \emph{observed integrated velocity dispersion} to distinguish from the intrinsic velocity dispersion ($\sigma$).

For massive galaxies, the central potential well is dominated by stars, therefore the \emph{intrinsic} stellar velocity dispersion ($\sigma_{\star}$) - either with or without projection effects - will be more representative than the ionized gas velocity dispersion ($\sigma_{g}$), which does not trace collisionless orbits and is unlikely to trace the same structures as the stellar component. The stellar velocity dispersions of galaxies, especially for the early-type subset, have been demonstrated to be tightly connected to their luminosities, stellar masses, and sizes \citep[e.g.][]{faberjackson, djorgovski:87, dressler:87} as well as the supermassive black holes at their centers \citep[e.g.][]{magorrian:98}. The number of galaxies as sorted by $\sigma_\star$ -- the velocity dispersion function -- and its evolution have been studied from a galaxy evolution perspective \citep[e.g.][]{sheth:03, bezanson:11, bezanson:12}, but is also important for understanding weak and strong lensing studies \citep[e.g.][]{chae:10,mason:15}.

\begin{figure*}[!t]
\centering
\includegraphics[width=0.9\textwidth]{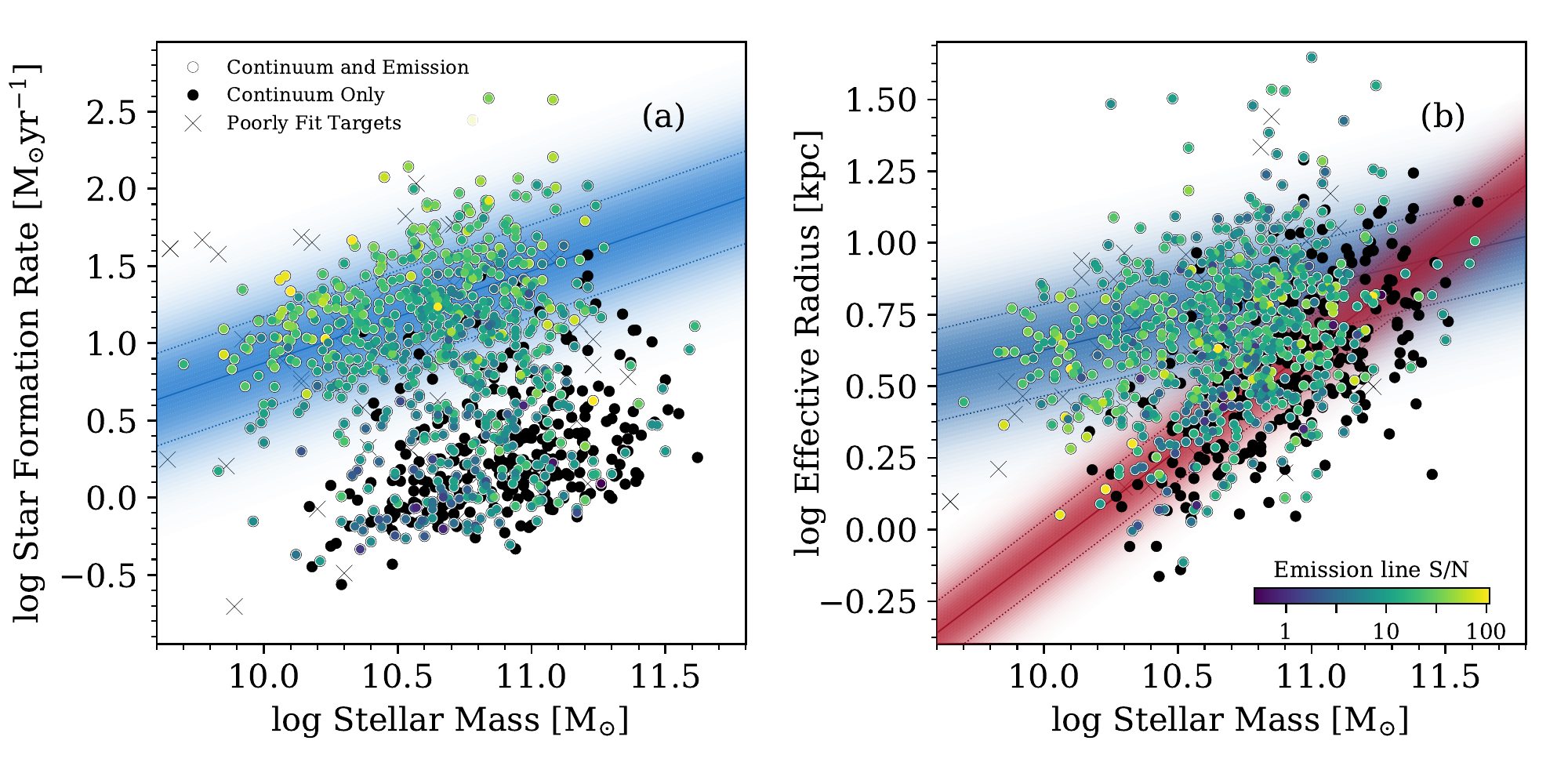}
\caption{The LEGA-C sample in star formation rate versus stellar mass with the \citet{whitaker:12b} relation (blue band, left panel) and in effective radius versus stellar mass and star-forming and quiescent relations from \citet{wel:14} (blue and red bands, right panel). All galaxies with successful dynamical fits are indicated by circles, where symbols are colored by the S/N of their brightest emission line. Targets with poorly fit spectra are included as crosses.}
\label{fig:select}
\end{figure*}

Measuring $\sigma'_{\star, int}$ requires high S/N in the continuum, in contrast with measuring $\sigma'_{g, int}$ from brighter emission lines. This renders measurements of this key parameter increasingly difficult as spectroscopic studies of galaxies push into the high redshift Universe. It would be beneficial to use the less expensive $\sigma'_{g, int}$ as a proxy for $\sigma'_{\star, int}$ - and ultimately $\sigma_{\star}$ - when emission lines are detected. \citet{ho:09} found that in the centers of local galaxies, the gas and stellar velocity dispersions were strongly correlated, but found trends with other galaxy properties in addition to a net offset $\langle\sigma'_{g, int}/\sigma'_{\star, int}\rangle=0.8$. Conversely \citet{chen:08} found very good average agreement between $\sigma'_{g, int}$ and $\sigma'_{\star, int}$ as measured from the \emph{spatially-integrated} spectra of emission line galaxies in the Sloan Digital Sky Survey (SDSS). 

Of additional concern is the fact that at low redshift, ionized gas emission is not ubiquitous and is uncommon in the most massive galaxies \citep[e.g.,][]{pandya:17}. However, at higher redshifts the population of massive galaxies includes a higher fraction of star-forming galaxies \citep[e.g.][]{muzzin:13,tomczak:14} and a greater diversity of spectroscopic properties \citep[e.g.][]{dokkum:11}. It follows that $\sigma'_{g, int}$ could be measured for an increasingly representative sample of galaxies at high-redshift, precisely where $\sigma'_{g, int}$ is the most valuable proxy.

In this Letter we test the relationship between stellar and gas 1D kinematics for the first statistical and representative sample of galaxies at significant lookback time. The sample of $\sim1000$ galaxies selected from the Large Early Galaxy Astrophysics Census (LEGA-C) provide the necessary deep continuum spectroscopy to measure $\sigma'_{\star, int}$ and sufficient demographic range to probe trends between ionized gas and stellar kinematics at $z\sim0.8$. We describe the dataset in \S \ref{sect:data}, explore the relationship between $\sigma'_{g, int}$ and $\sigma'_{\star, int}$ in \S \ref{sect:sig_sig}, and discuss the implications of our findings in \S \ref{sect:disc}. Throughout we assume concordance cosmology $\Omega_{\lambda}=0.7, \Omega_{M}=0.3$, and $H_0=70\mathrm{km s^{-1} Mpc^{-1}}$.

\section{Data}\label{sect:data}
\begin{figure*}
\centering
\includegraphics[width=0.95\textwidth]{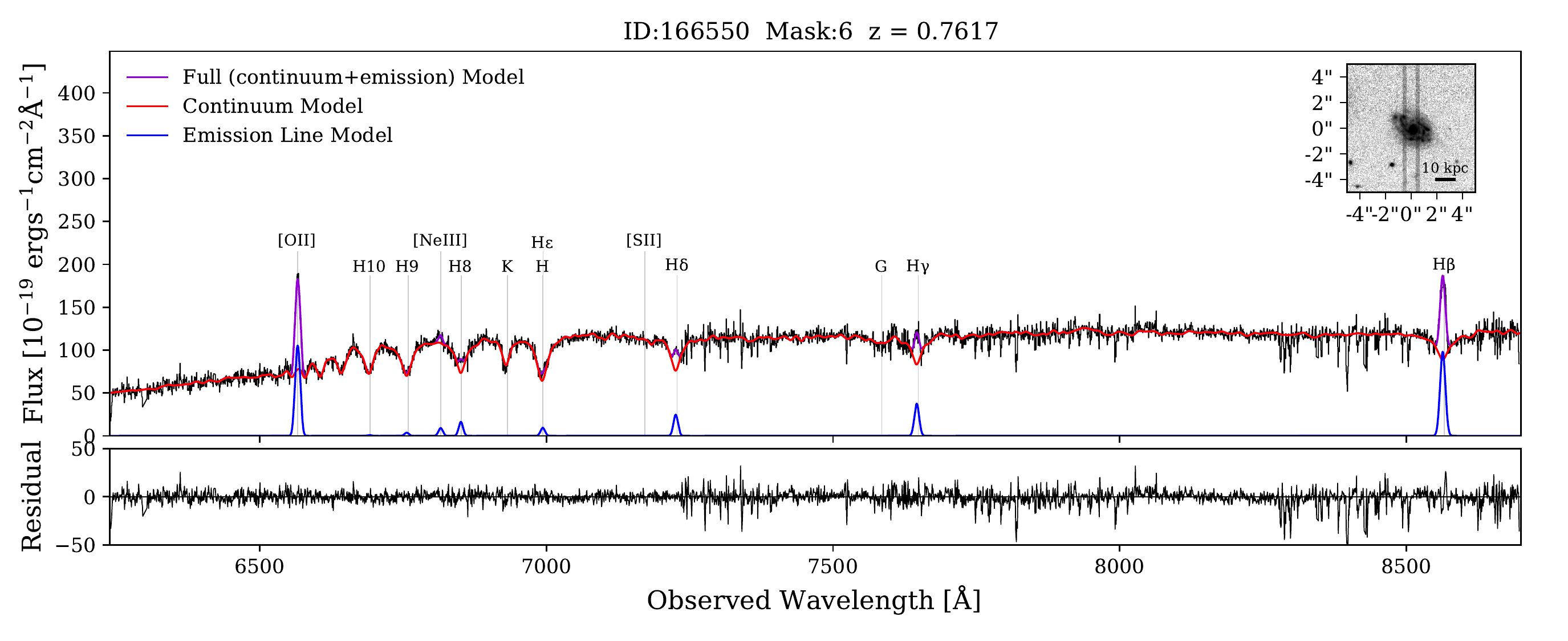}
\includegraphics[width=0.95\textwidth]{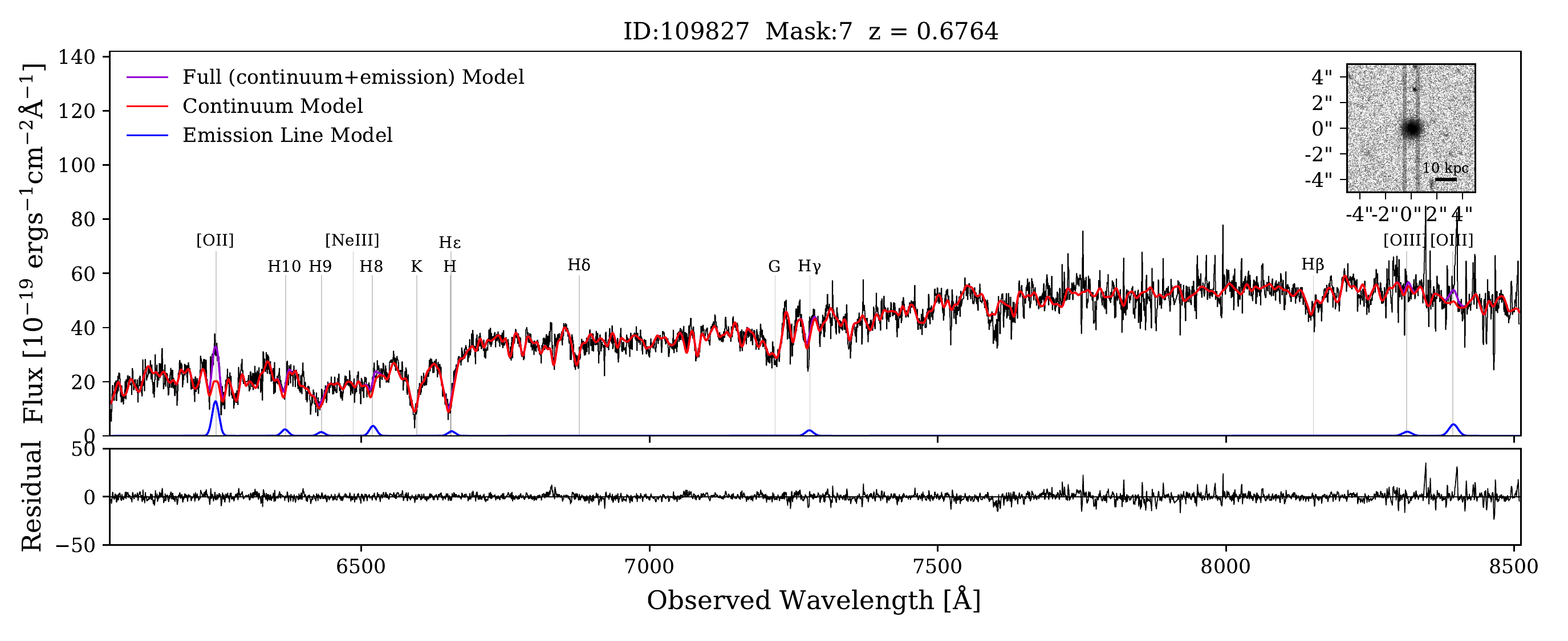}
\caption{LEGA-C spectra of a star-forming (ID:166550, top) and quiescent galaxy (ID:109827, bottom). All galaxies are fit with a combination of stellar population templates (red) and gaussian emission lines (blue) to model ionized gas lines (with combined fit, purple). The ACS F814W images with vertical LEGA-C slit are shown in right insets.}
\label{fig:spectra}
\end{figure*}

\subsection{The Large Early Galaxy Astrophysics Census}
This paper is primarily based on data release 2 (DR2) of the LEGA-C survey, an ESO Large Spectroscopic Program including ultradeep spectroscopy of $\sim3000$ massive $0.6<z<1.0$ galaxies in the COSMOS field using VIMOS on the VLT.  For a more detailed description of the survey, data reduction, and quality we refer to \cite{wel:16,straatman:18}; we briefly summarize here. This paper is based on primary targets selected using a redshift-dependent magnitude limit, $K_{AB}=20.7-7.5\log((1+z)/1.8)$, which yields a mass-complete sample above $\log(M_{\star}/M_{\odot})\gtrsim10.3$. Observations were taken using the HRred grating, which produces $R\sim2500$ spectra over $\sim6300-8800\mathrm{\AA}$, with a dispersion of $0.6\mathrm{\AA\,pix^{-1}}$. {The 1D spectra are extracted from the 2D spectra with a Moffat kernel in the spatial direction that varies from galaxy to galaxy but typically has a FWHM of 1" (see Straatman et al. 2018 for details). We do not expect this to significantly impact the measured velocity dispersions as observed velocity dispersion profiles are very flat within this window, largely due to beam smearing \citep{bezanson:18a}.  We note that this aperture is comparable to the galaxy sizes, unlike similar measurements for local galaxies.} Each mask includes roughly 100 primary targets and is observed for $\sim$20\,hours, reaching {a typical continuum $S/N\sim20{\rm\AA^{-1}}$ in the observed frame 1D extracted spectra.}

\subsection{Ancillary Data}
The LEGA-C sample is selected from the v4.1 \citet{muzzin:13ultravista} Ks-selected catalogs in the UltraVista/COSMOS field, which provides a wealth of ancillary data, including PSF-matched photometry in 30 bands from 0.15-24$\mu m$ from a number of facilities including GALEX \citep{martin:05}, CFHT/Subaru \citep{capak:07}, UltraVISTA \citep{mccracken:12}, and S-COSMOS \citep{sanders:07}. Stellar masses are measured as in the \citet{muzzin:13ultravista} data release, but with fixed spectroscopic redshift, using the Fitting and Assessment of Synthetic Templates (FAST) code \citep{kriek:09}, assuming \citet{bc:03} templates, a \citet{chabrier:03} initial mass function, exponentially declining star formation histories, and a \citet{calzetti:00} dust law. ``UV+IR'' Star formation rates (SFR) are measured from the UV fluxes plus reradiated dust emission measured from the MIPS 24$\mu m$ flux following \citet{whitaker:12b}. 

Fig.\,\ref{fig:select}a shows the SFR versus stellar mass for galaxies in the primary LEGA-C DR2 sample ($\mathrm{use}=1$, see \citet{straatman:18} for details) with the \citet{whitaker:12b} relation (blue line and band). Colored and black symbols indicate galaxies with and without emission lines in their LEGA-C spectra, with color corresponding to the S/N of the brightest observed emission line. Black crosses identify 176 galaxies for which spectral modeling failed (either flagged visually, {for example in the case of obvious broad line AGN,} or with $\geq20\%$ uncertainties in $\sigma'_{\star, int}$).  We note that galaxies with emission lines are not confined to the locus of star-forming galaxies (e.g., as identified photometrically); the LEGA-C spectra uncover a significant population of quiescent galaxies with emission lines \citep[see also][]{straatman:18}. 

Morphologies are measured from the COSMOS HST/ACS F814W v.2.0 mosaic \citep{cosmosacs,massey:10}. Each galaxy is modeled by a single S\'ersic profile following \citet{wel:12}, which produces best-fit effective radii, S\'ersic indices, axis ratios, and position angles. Fig.\,\ref{fig:select}b shows effective radius (semi-major, rest-frame 5000\AA) versus stellar mass for the sample. \citet{wel:14} relations and scatter for star-forming and quiescent galaxies at $z\sim0.75$ are indicated by blue and red bands.

Catalogs of X-ray detections in the COSMOS field exist based on data collected by the XMM and Chandra telescopes. The XMM-COSMOS survey provides the XMM Point-like Source Catalog \citep{cappulluti:09} in three bands (0.5-2.0\,keV, 2.0-4.5\,keV, and 4.5-10.0\,keV to $7.27\times10^{-16}$, $4.96\times10^{-15}$, and $8.2\times10^{-15}\,\rm{erg\,cm^{-2}\,s^{-1}}$ depths) and the Chandra COSMOS Legacy 4.6 Ms survey also includes three bands of X-ray fluxes (0.5-2.0\,keV, 2.0-5.0\,keV, and 5.0-10.0\,keV to $2.2\times10^{-16}$, $1.5\times10^{-15}$, and $8.9\times10^{-16} \rm{erg\,cm^{-2}\,s^{-1}}$ depths) \citep{civano:16,marchesi:16}. We match the LEGA-C catalog to 96 XMM-COSMOS and C-COSMOS sources within a 1'' radius. 

\subsection{Stellar and Ionized Gas 1D Kinematics}
The observed gas and stellar velocity dispersions ($\sigma'_{g, int}$ and $\sigma'_{\star, int}$) are measured for every galaxy as the gaussian line width (for emission lines) and broadening (for the stellar continuum) in the optimally extracted 1D spectra using Penalized Pixel-Fitting (pPXF) \citep{cappellari:04,cappellari:17}. The software models each galaxy spectrum as a combination of one or more stellar population templates and emission lines (at instrumental resolution) convolved with a {single} gaussian broadening as well as multiplicative and additive polynomials to account for uncertainties in the overall spectral shape. We adopt a third order multiplicative polynomial and an additive polynomial with one degree of freedom per 100$\mathrm{\AA}$ , however we verify that due to the extremely high S/N nature of these spectra the fits are largely insensitive to polynomial choice with $\geq2$nd order multiplicative polynomial. The continuum is modeled with high resolution (R=10,000) theoretical single stellar population templates.  These templates were produced with the FSPS package \citep{conroy:09}, using an unpublished grid of theoretical spectra computing using the ATLAS12/SYNTHE routines \citep{kurucz:11}; see \citet{conroy:12} for details. We verify that these agree with fits using \citet{vazdekis:99} lower-resolution empirical templates for $\sigma'_\star\gtrsim$100\,$\mathrm{km\,s^{-1}}$ with a scatter of $\lesssim\,$7\%. {The observed stellar velocity dispersions are fixed to the same value for all stellar templates and the fit is luminosity-weighted. Although some of the stronger features (Balmer lines) will be dominated by younger stars, this wavelength range also contains a wealth of weaker metal lines (see Figure \ref{fig:spectra}) that are sensitive to older stellar populations.}

Examples of LEGA-C spectra with best-fitting models are shown in Fig.\,\ref{fig:spectra}. Gas velocity dispersions are measured from a combination of emission lines ([NeV],\,[NeVI],\,H10,\,H9,\,H8,\,H$\epsilon$, H$\delta$,\,H$\gamma$, H$\beta$, [OII] doublet, [NeIII], and [OIII], depending on the wavelength coverage). The individual line normalization is free, but the $\sigma'_{g, int}$ is the same for all lines. Emission and absorption templates are fit simultaneously{, starting at the instrumental resolution and broadening and normalizing to fit the spectra. A} fit is accepted if the emission lines contribute at least 25\% of the flux in one part of the spectrum. If this is not the case, or if the redshifts of the stellar and gas templates differ by too much ($|z_{gas}-z_{stars}|>0.003$) we refit with stellar templates alone. We visually inspect all fits to verify that this process correctly identifies emission lines. {Almost all galaxies have integrated velocity dispersions that far exceed the instrumental resolution.} Uncertainties in $\sigma'$ are based on formal uncertainties and are rescaled based on duplicate observations of individual targets \citep[see][]{straatman:18}. 

\section{Comparison between Integrated Stellar and Ionized Gas Kinematics}\label{sect:sig_sig}

\begin{figure}
\centering
\includegraphics[width=0.45\textwidth]{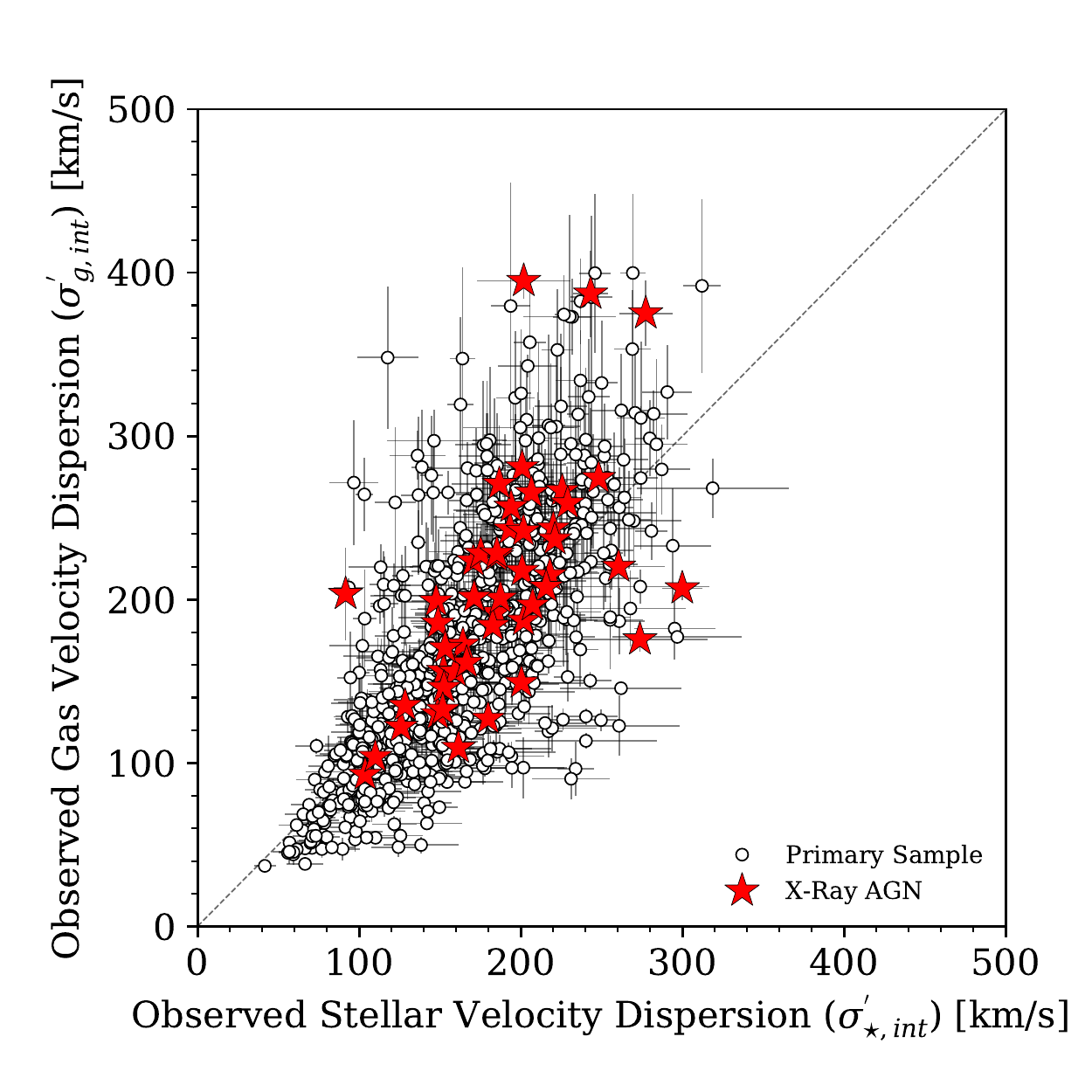}
\caption{Gas versus stellar observed velocity dispersion measurements for galaxies with detected emission lines in the LEGA-C survey.  Galaxies are indicated by open circles and 49 galaxies with X-ray detections are highlighted by red stars. The two measures of galaxy kinematics agree for the population, with significant scatter (overall 0.13\,dex).  X-ray AGN, for which emission line widths are likely to be sensitive to the central engine in addition to the galaxy potential, are indeed offset to higher $\sigma'_{g, int}$ than $\sigma'_{\star, int}$ but do not account for all outliers.}
\label{fig:sigma_sigma}
\end{figure}

In this section we compare observed velocity dispersions from stars and ionized gas in 813 galaxies with detected emission lines and reliable spectral fits (with $<20\%$ errors on $\sigma'_{g, int}$ and $\sigma'_{\star, int}$). Fig.\,\ref{fig:sigma_sigma} shows the $\sigma'_{g, int}$ versus $\sigma'_{\star, int}$. Overall the two measures of the galaxy kinematics agree ($\langle\log(\sigma'_{g, int}/\sigma'_{\star, int})\rangle$=$-0.003$\,dex), but with significant scatter (0.13\,dex, with 0.05\,dex due to observational errors). 

There are many reasons to expect differences between gas and stellar kinematics. Stellar kinematics are most sensitive to the distribution of mass in the inner parts of galaxies where stars dominate, whereas the ionized gas can have a range of spatial distributions. Furthermore, emission line kinematics are influenced by gas inflows and outflows and central AGN activity. Galaxies with X-ray detections, likely AGN hosts, are indicated by red stars and generally lie at elevated $\sigma'_{g, int}$. However, these galaxies by no means account for all elevated $\sigma'_{g, int}$ measurements and there may be weaker AGN that are not detected in X-ray. Furthermore, there is a significant subset of galaxies with broader stellar than gas kinematics (25 with $(\sigma'_{\star, int}/\sigma'_{g, int})>2$). For these galaxies the ionized gas is not probing the full galaxy potential well and $\sigma'_{g, int}$ would significantly underestimate dynamical masses, however we note that the uncertainties on these {low}-$\sigma'_{g, int}$ measurements are often large.

\begin{figure*}[t]
\centering
\includegraphics[width=\textwidth]{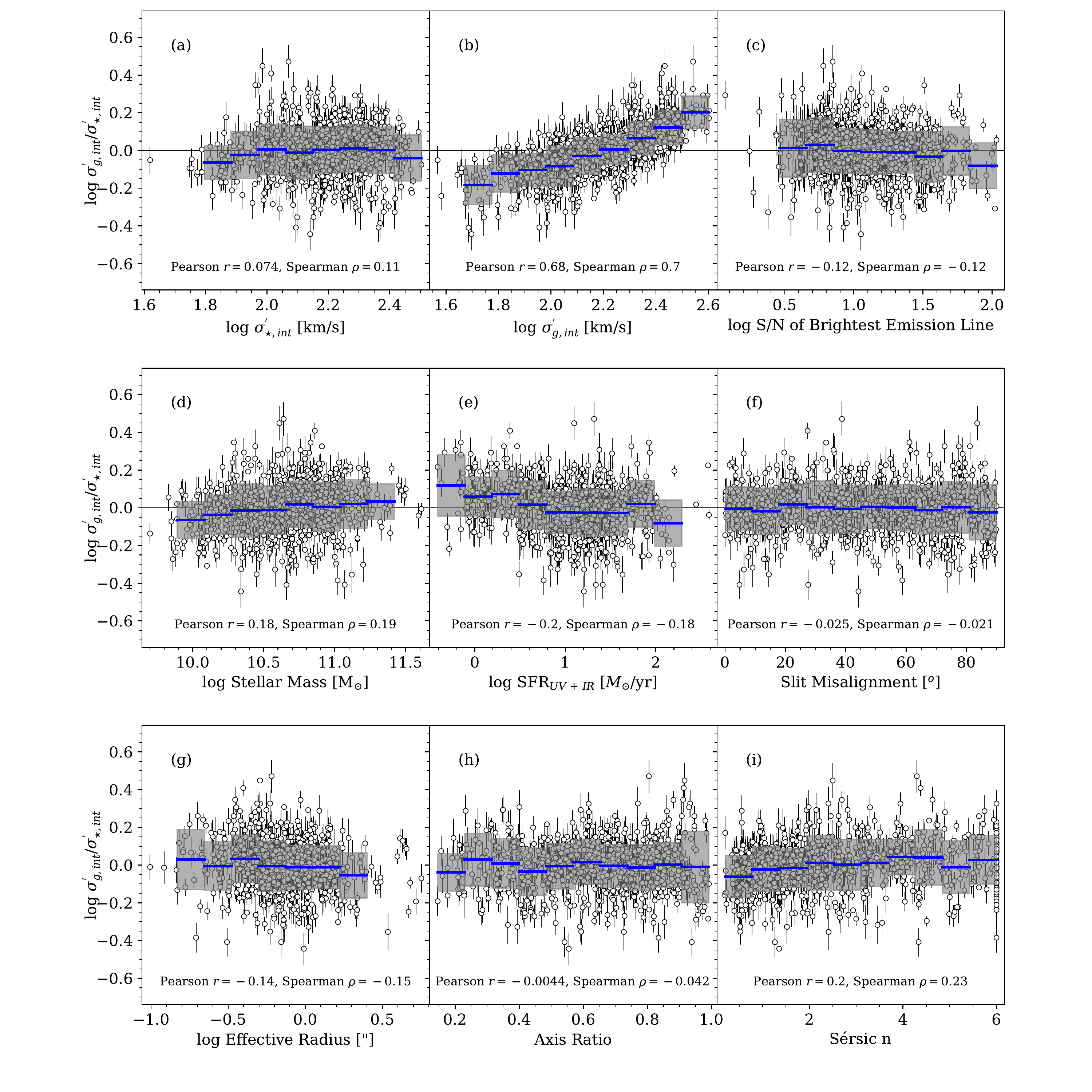}
\caption{Ratios between observed velocity dispersions as a function of $\sigma'_{\star, int}$ and $\sigma'_{g, int}$ (a and b), maximum emission line S/N (c), stellar populations (d and e), slit alignment (f), and structural (g - i) properties. Blue lines and grey bands indicate running average and scatter. Pearson and Spearman correlation coefficients are noted in each panel.}
\label{fig:resids}
\end{figure*}

We find that the scatter between $\sigma'_{g, int}$ and $\sigma'_{\star, int}$ persists for all subsets of galaxies. In Fig.\,\ref{fig:resids} we explore residuals between gas and stellar $\sigma'$'s with either observed velocity dispersion and S/N (top row), stellar populations and slit alignment (middle row), and galaxy structures (bottom row). Fig.\,\ref{fig:resids}a indicates that the ratio of $\sigma'_{g, int}/\sigma'_{\star, int}$ is roughly independent of $\sigma'_{\star, int}$. However, there is a strong trend (Fig.\,\ref{fig:resids}b) with $\sigma'_{g, int}$ such that low values ($\sigma'_{g, int}\lesssim100\mathrm{km\,s^{-1}}$) will underestimate $\sigma'_{\star, int}$, and the opposite for high observed gas velocity dispersion ($\sigma'_{g, int}\gtrsim200\mathrm{km\,s^{-1}}$). At low $\sigma'_{g, int}$, this trend is likely imposed by the K-band selection of the LEGA-C survey, which preferentially excludes galaxies with low $\sigma'_{\star, int}$. This is the only strong trend (see correlation coefficients) in scatter or residuals between $\sigma'_{g, int}$ and $\sigma'_{\star, int}$. Fig.\,\ref{fig:resids}c, shows the residuals versus stellar mass and indicates that scatter decreases slightly at the lowest and highest stellar masses. However, emission line occurrence rate also decreases at the highest masses. The scatter between $\sigma'$s is roughly constant with UV+IR SFR (Fig.\,\ref{fig:resids}d) and we find no trends with slit misalignment with respect to photometric semi-major axis (Fig.\,\ref{fig:resids}e). One might expect that the correspondence between gas and stellar kinematics would correlate more strongly with stellar structures, however we do not find evidence for this in the bottom row of Fig.\,\ref{fig:resids} (size, axis ratio, and S\'ersic index). In Fig.\,\ref{fig:resids}h the scatter between $\sigma'$s increases somewhat for the roundest ($b/a\gtrsim0.8$) galaxies. The roundest galaxies may exhibit scatter between $\sigma'$s because they probe different disk versus bulge morphologies.  Fig.\,\ref{fig:resids}i shows the scatter between $\sigma'_{g, int}$ and $\sigma'_{\star, int}$ is similar for galaxies of all profile shapes. In all bins of galaxy structural and stellar populations, except at the lowest masses ($\log M_{\star}/M_{\odot}<10$) and very large sizes ($r_e\gtrsim12$kpc), the average $\log \sigma'_{g, int}/\sigma'_{\star, int}$ is less than $\sim$0.1\,dex.

\begin{figure*}[!t]
\centering
\includegraphics[width=0.66\textwidth]{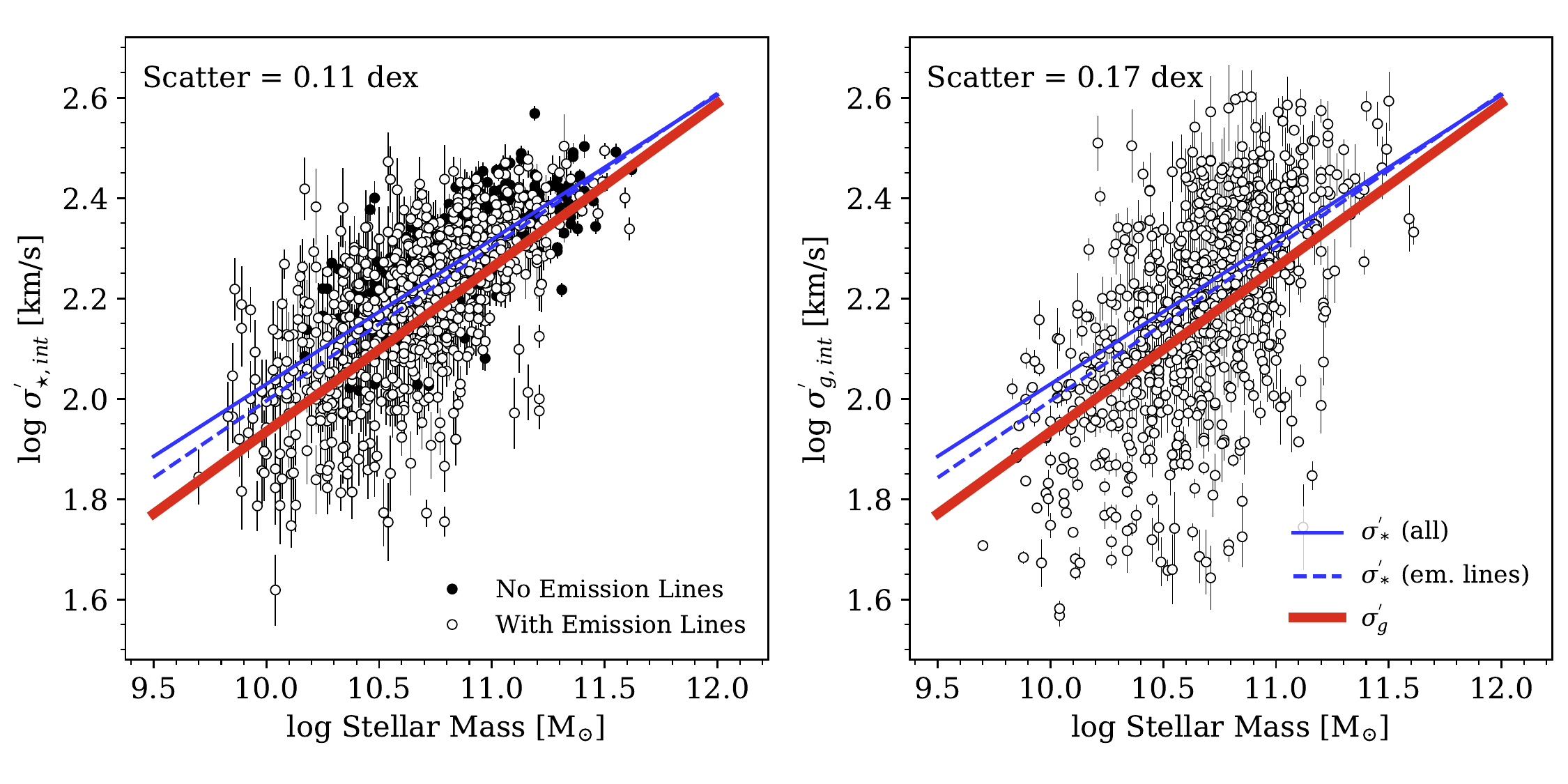}
\includegraphics[width=0.33\textwidth]{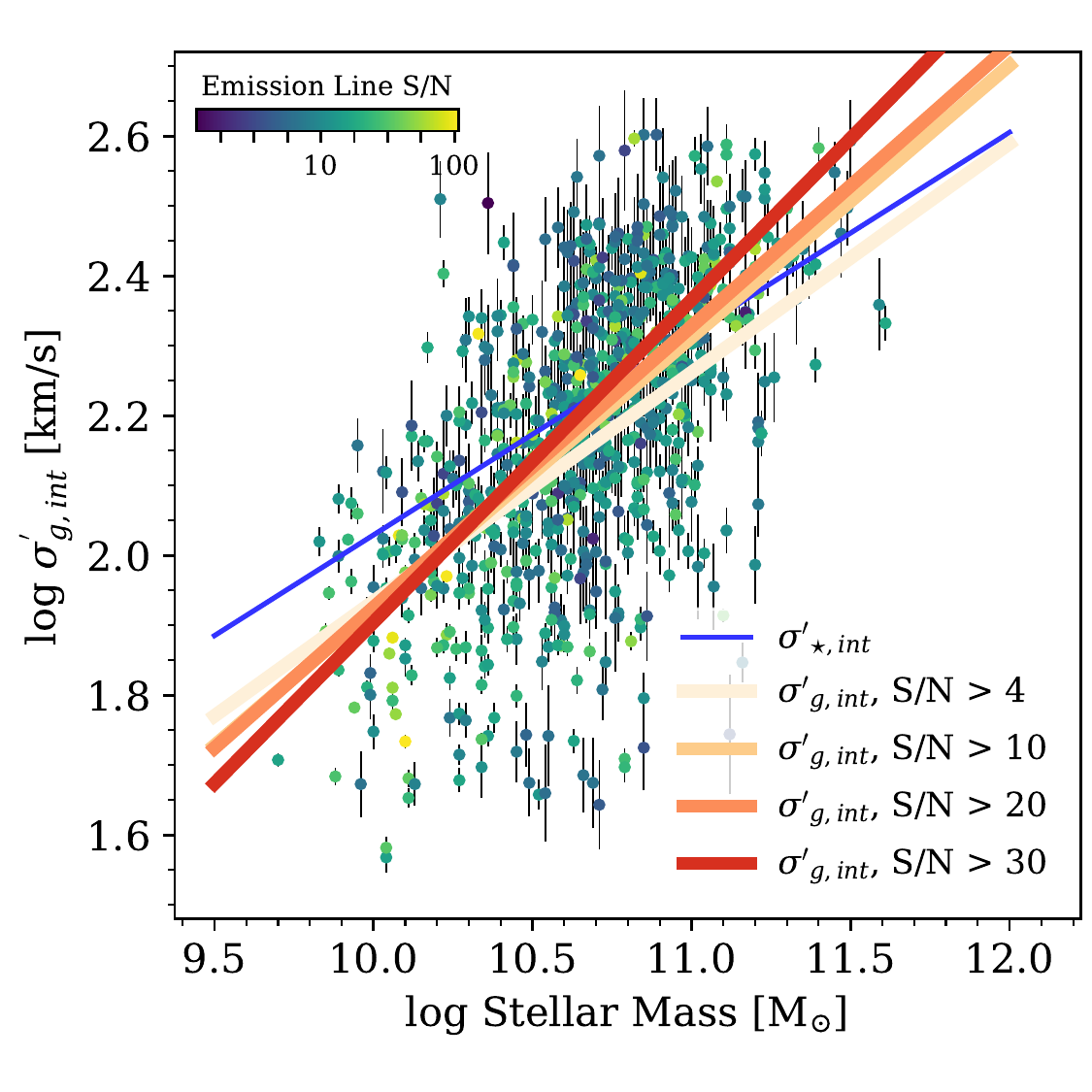}
\caption{Mass Faber-Jackson relation measured from absorption ($\sigma'_{\star, int}$, left panel) and emission ($\sigma'_{g, int}$, center and right panels). Open/colored and black symbols indicate galaxies with and without emission lines. Best-fit linear relations with $\sigma'_{\star, int}$ (solid line for all galaxies and dashed for only those with emission lines)  and $\sigma'_{g, int}$ (thick {red} line) are shown in the left and center panels. The right panel reproduces the center panel with symbols colored by emission line S/N along with best-fit relations for subsamples above a range of S/N thresholds ({red and orange} lines). The Mass FJ relations are very similar for the full population, indicating that either measure can be used to estimate galaxy dynamics, however the $\sigma'_{\star, int}$ relation is much tighter and somewhat shallower, with a vertical scatter of 0.11\,dex relative to 0.17\,dex measured from $\sigma'_{g, int}$. This increased scatter translates to an additional uncertainty of $\sim0.2\,\rm{dex}$ in dynamical mass. Furthermore, relying on bright emission features will bias towards an increasingly steep relation.}
\label{fig:mass_fj}
\end{figure*}

Ultimately one would like to use velocity dispersions, measured from either $\sigma'_{g, int}$ or $\sigma'_{\star, int}$, to probe scaling relations and estimate dynamical masses. In Fig.\,\ref{fig:mass_fj} we show $\sigma'_{\star, int}$ (left) and $\sigma'_{g, int}$ (center and right) versus stellar masses or the ``stellar mass'' Faber-Jackson (mass FJ) relation \citep{faberjackson}. We note that for $\sigma'_{g, int}$, which is related to the circular velocity of a disk modulo beam-smearing and projection effects, this is related to the {modified} \citet{tullyfisher} relation \citep[e.g., {$S0.5$ - $M_{\star}$ relation,}][]{kassin:07, straatman:17}. We refer the reader to Straatman et al.~in prep, for further 2D analysis of these observational effects. 
We expect both of these relations to have some intrinsic scatter, which for the FJ relation correlates with galaxy size - corresponding to the Fundamental Plane \citep{djorgovski:87,dressler:87}. We fit linear relations using a least-squares fitting algorithm and estimate the uncertainty with a 1000-realization bootstrap analysis. Using $\sigma'_{\star, int}$ for the full sample (solid blue lines in Fig.\,\ref{fig:mass_fj}) we find:
\begin{equation}
\log \sigma'_{\star, int}=(-0.85\pm0.11)+(0.29\pm0.01)\log\,M_{\star},
\end{equation}

\noindent which is consistent within $1\sigma$ with results from a smaller sample at $z\sim0.7$ \citep{bezanson:15}. We also fit the subsample of galaxies with emission lines and find:

\begin{equation}
\log \sigma'_{\star, int}=(-1.07\pm0.15)+(0.31\pm0.01)\log\,M_{\star}, 
\end{equation}

\noindent (dashed blue lines).These relations are consistent at the 95\% confidence level, however the slight tension between the fits emphasizes that selecting galaxies with emission lines biases the sample and impacts the measured scaling relations. Finally, we fit the mass FJ relation using $\sigma'_{g, int}$ (thick {red} line) and find:
\begin{equation}
\log \sigma'_{g, int}=(-1.34\pm0.44)+(0.33\pm0.04)\log\,M_{\star},
\end{equation}
\noindent which is consistent within $1\sigma$ with other fits, with higher uncertainties. We note that this relation is steeper than for the full sample or from $\sigma'_{\star, int}$, a bias which increases when only galaxies with the strongest emission lines are used ({red and orange} lines in right panel). {Neither $\sigma_{int}^{'}$ is strongly correlated with either S/N of the strongest emission line or in the continuum (e.g. measured at rest-frame 4000 \,{\AA}). However, note that below below $S/N\sim4$, $\sigma_{g,int}^{'}$ values are likely biased high, therefore we exclude galaxies with the lowest S/N emission lines} The measured relation becomes progressively steeper with stronger emission lines.

Although there is general agreement amongst these relations, the observed scatter is significantly higher when $\sigma'_{g, int}$ is used (0.17\,dex with $
\sim$0.06\,dex due to observational errors in $\sigma'_{g, int}$) than for stellar $\sigma'$ (0.11\,dex with 0.04\,dex due to errors). This difference in quadrature between the scatter in the two relations ($0.13$dex) is equal to the scatter between $\sigma'_{g, int}$ and $\sigma'_{star}$. Although population-averaged scaling relations can be approximated from emission line kinematics, the slopes may be biased and the scatter about those relations will be significantly overestimated. Since typical spectroscopic surveys are shallower than LEGA-C, this suggests that previously published scalings between mass and {1D} emission line width {\citep[e.g.][]{mocz:12}} may have biased slopes. 

{We emphasize that this analysis is solely based on 1D kinematics, which could easily be significantly sensitive to inclination effects. This is particularly true for star forming galaxies, which are largely rotationally supported at these redshifts \citep[e.g.,][]{kassin:07}. We test this by comparing the comparing $\sigma_{g,int}^{'}$ and $\sigma'_{\star, int}$ with their averages evaluated in different mass bins for star forming and quiescent galaxies as a function of axis ratio. We find that at fixed stellar mass both $\sigma_{g,int}^{'}$ and $\sigma'_{\star, int}$ can be up to $\sim$0.1-0.3 dex below the average $\sigma_{int}^{'}$ for the roundest ($b/a>0.8$) star-forming galaxies, whereas for all other axis ratios and all quiescent galaxies, the agreement is very good on average. This effect generally impacts both gas and stellar observed velocity dispersion similarly; both $\sigma_{g,int}^{'}$ and $\sigma'_{\star, int}$ are offset from their average values coherently. This may explain the lack of residuals in Figure \ref{fig:resids}h; both gas and stellar dynamics are similarly poor tracers of the dynamical mass for round star-forming galaxies. We verify that excluding round galaxies from the fits presented in Figure \ref{fig:mass_fj} yields relations that are consistent with Equations 1-3. The scatter is only 0.095dex about the $\sigma'_{\star, int}$ mass Faber-Jackson relation and 0.15dex about the relation derived using $\sigma'_{g, int}$, implying that inclination effects likely contributes $\sim$0.06dex and 0.08dex respectively.}

\section{Discussion and Conclusions}\label{sect:disc}

The extraordinary high S/N spectroscopy from the LEGA-C survey opens up a new window into the stellar continuum of massive galaxies at cosmological distances, while the magnitude-limited survey design facilitates an investigation of trends within the galaxy population. In contrast, most spectroscopic surveys of high-z galaxy kinematics are limited by depth and/or resolution to emission line studies. This combination has facilitated the comparison between stellar and ionized gas 1D kinematics in 813 massive galaxies at $z\sim0.8$. 

We emphasize that there is significant scatter between $\sigma'_{\star, int}$ and $\sigma'_{g, int}$ for galaxies of all structures and stellar populations. Overall the $0.13$\,dex scatter is slightly lower for galaxies with intermediate axis ratios, S\'ersic indices, and the highest masses, but in all cases is $\gtrsim0.1$\,dex. Although this may seem like a small price to pay to rely on emission line spectroscopy, we caution that for any individual galaxy this translates to an intrinsic uncertainty of $\sim0.12$\,dex on $\sigma'$ when observed velocity dispersion is measured from emission lines. This uncertainty propagates to an uncertainty of 0.24\,dex on dynamical mass, on top of the other measurement errors and systematic uncertainties such as conversion between $\sigma'_{\star, int}$ and intrinsic $\sigma_{\star}$ or the Virial constants. This uncertainty is comparable to systematic uncertainties in the stellar masses of high redshift galaxies \citep[e.g.][]{muzzin:09}.

We are unable to identify a specific population of galaxies for which scatter between $\sigma'$s varies dramatically or the two measures are systematically offset, although we find a bias at fixed $\sigma'_{g, int}$. It is easy to imagine selecting a galaxy population, particularly at high redshift, that also happens to have significantly discrepant stellar and ionized gas distributions, which leads to differences between $\sigma'_{g, int}$ and the intrinsic or stellar $\sigma'$ for the full population. This could be at play, for example, in the compact star-forming galaxies observed at $z\sim2$ to have systematically lower $\sigma'_{g, int}$ than expected from their high stellar masses and small effective radii \citep[e.g.][]{barro:14, dokkum:15, barro:16}. Given the 0.13\,dex scatter, it would be easy to account for a factor of 1.5-2 in converting $\sigma'_{g, int}$ to $\sigma_{\star,predicted}$, which is sufficient e.g., to measure dynamical masses for all but the most extreme few galaxies. 
The good agreement between stellar and gas observed velocity dispersions implies that the less observationally expensive quantity, $\sigma'_{g, int}$, can be used to measure overall scaling relations for emission line galaxies. Interestingly, we find that the stellar mass Faber-Jackson relation is similar for the full population of massive galaxies, measured from $\sigma'_{\star, int}$, and for galaxies with emission lines from $\sigma'_{g, int}$, with a slight bias towards a steeper slope that increases as samples are limited to more prominent emission lines. Furthermore the uncertainty in the relation measured with $\sigma'_{g, int}$ and the observed scatter is significantly increased.  Although it would be tempting, for example, to measuring the velocity dispersion function using $\sigma'_{g, int}$, one must be extremely careful in accounting for the observed bias in emission line selections.

\acknowledgements
The authors would like to thank the anonymous referee for their constructive feedback that improved this publication. This research made use of Astropy \citep{astropy}. Based on observations collected at the European Organisation for Astronomical Research in the Southern Hemisphere under ESO programme 194.A-2005.

\end{document}